\documentclass[12pt,preprint]{aastex}
\newcommand{\twcooz}{$^{12}$CO $J$=1$-$0}
\newcommand{\twco}{$^{12}$CO}

\newcommand{\thco}{$^{13}$CO}

\newcommand{\etal}{et al.}
\newcommand{\eg}{e.g.}

\newcommand{\h}{$^{\rm h}$}
\newcommand{\m}{$^{\rm m}$}
\newcommand{\s}{$^{\rm s}$}
\newcommand{\kms}{km s$^{-1}$}
\newcommand{\jybm}{Jy bm$^{-1}$}
\newcommand{\jykms}{Jy km s$^{-1}$}
\newcommand{\jybmkms}{Jy bm$^{-1}$ km s$^{-1}$}

\newcommand{\msun}{M$_\odot$}

\shorttitle{CO in HCG 92}
\shortauthors{Petitpas \& Taylor}

\begin{document}

\title{A High Resolution Mosaic of Molecular Gas in Stephan's Quintet}

\author{Glen R. Petitpas}
\affil{Smithsonian Astrophysical Observatory}
\affil{645 N. A'ohoku Place, Hilo, HI, USA 96720}
\email{gpetitpas@cfa.harvard.edu}
\and
\author{Christopher L. Taylor}
\affil{California State University, Sacramento}
\affil{6000 J Street, Sacramento, CA, USA 95819}
\email{ctaylor@csus.edu}

\begin{abstract}

We present high resolution \twcooz\ observations of the molecular gas
in the Hickson Compact Group Stephan's Quintet (HCG92). Our
observations consist of multiple pointing and mosaics covering all the
regions where CO and star formation has been detected. Within the
100\arcsec\ field of view centered on the eastern-most tidal tail, we
detect three clumps of emission that may be partially resolved at our
resolution of 8\arcsec; two of these are new detections not previously
seen in ISM studies of this region. Two of these clumps lie in the
optical tidal tail, while the third lies to the southeast and is
coincident with a large \ion{H}{1} feature, but does not correspond to
any features at other wavelengths. We also tentatively detect CO
emission from the star forming regions in the ``Old Tail''
corresponding to recent star formation activity detected in recent UV
and H$\alpha$ observations. Observations of the rest of the compact
group do not show detections even though strong emission was detected
with single dish telescopes, which suggests the CO emission originates
from a diffuse molecular gas cloud or from more at least three
separate clumps with separations of greater than around 3 kpc.

\end{abstract}


\section{Introduction \label{intro}}

The parallel studies of dwarf galaxies and the outer regions of spiral
galaxies have come together as awareness has grown of the apparent
formation of new dwarf galaxies in the tidal arms of merging and
interacting systems. These galaxies are referred to as ``Tidal Dwarf
Galaxies'' (hereafter TDGs). The concept of material from interacting
galaxies being thrown out of interacting galaxies and forming new
galaxies is old \citep{zwi56,sch78}. This idea has been revived since
the early 1990's by Mirabel and collaborators, who found patches of
enhanced optical emission along the tidal tails of interacting systems
\cite[\eg,][]{mir91} and subsequently found that these patches
correlate closely with local enhancements in the column density of
\ion{H}{1} \citep[\eg,][]{duc94,duc97}.

Studies of the physical properties of TDGs show that they are very
similar in gas mass and size to typical gas rich dwarfs such as dwarf
irregulars (dIs) and blue compact dwarfs
\citep[BCDs;][]{duc01,bra01}. One difference is that the metallicities
of TDGs tend to cluster at the metal rich end of the range found for
the general dI and BCD populations. In fact, the metallicities of TDGs
are consistent with their ISMs being composed of gas from the outer
disks of spiral galaxies.

Some TDGs and candidate TDGs show signs of recent star formation, such
as blue colors and H$\alpha$ emission \citep[\eg,][]{duc94,igl01},
which implies the presence of molecular gas. Because TDGs are more
metal rich than typical dIs and BCDs, they may be brighter in CO
emission than dIs and BCDs \citep{tay98}. Indeed, several TDGs have
been detected in CO emission \citep{bra01}, although not yet in
sufficient numbers to say for certain that they are easier to detect
than dIs and BCDs in CO.

Several issues surrounding TDGs remain unresolved, including whether
or not they are truly self-gravitating, newly formed galaxies, whether
the CO-to-H$_2$ conversion factor is higher than in dIs and BCDs, and
whether their molecular gas is formed {\it in situ} from atomic gas,
or is already in molecular form when expelled from the parent merging
system. To shed some light on these issues, we have made a wide-field
mosaic image in the \twcooz\ line of the Hickson Compact Group 92,
also known as Stephan's Quintet.

Stephan's Quintet is a group of five galaxies (only three are
interacting) contained in a relatively small region of space. The
result of living within such close proximity is that there are a
larger number of interactions between neighboring galaxies than is
found in larger clusters of galaxies, where the relative distances and
velocities are larger. It is believed that an intruder galaxy (NGC
7320c) passed through the center of this system, depositing a
reservoir of molecular gas that was later disrupted by the passage of
NGC 7318b to trigger star formation in the intergalactic medium
\citep[\eg,][]{cha04}.

This system is also the host of a recently discovered extragalactic
starburst \citep{xu99}, and has been mapped in CO at low spatial
resolution over most of its spatial extent \citep{lis02}, or at high
spatial resolution in a single field
\citep{gao00,lis04}. \citet{igl01} identified seven extragalactic
H$\alpha$ emitting regions in HCG 92 which may be TDGs. Ours is the
first wide-field, high resolution survey of CO emission in the compact
group, and in 6 of these seven TDG candidates.

\section{Observations and Reduction}

\subsection{BIMA Observations of \twcooz}

\begin{figure}
\includegraphics[angle=-90,scale=.8]{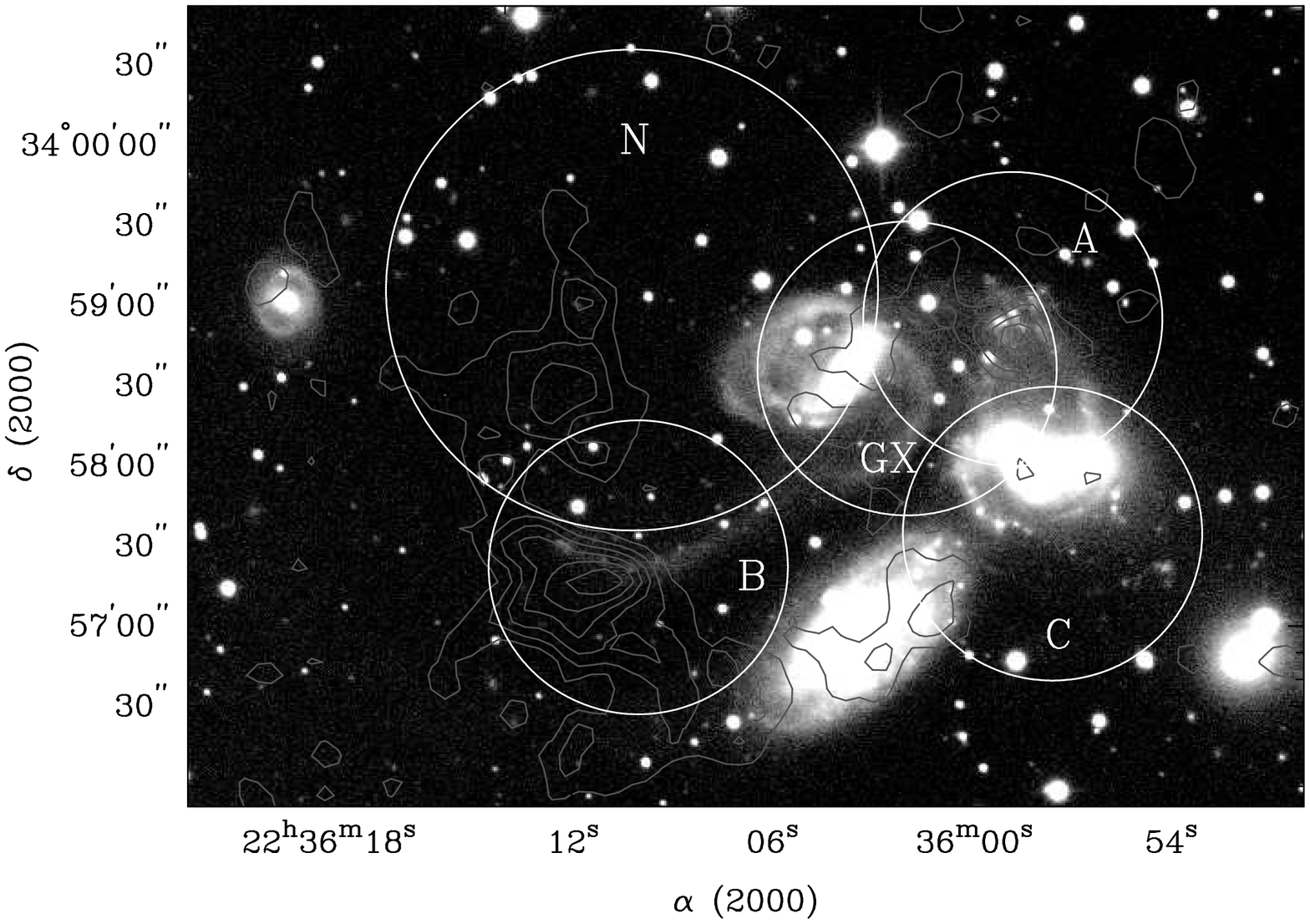}
\caption[f1.eps]{The four regions of HCG 92 observed by us with
the BIMA array (labeled A, B, C, and N) as well as the region
observed by \citet{gao00} (labeled GX). The underlying image is an
NOAO photograph taken by N.A.Sharp/NOAO/AURA/NSF. The gray contours
are the \ion{H}{1} data from \citet{wil02}.
\label{schematic}
}
\end{figure}

We have observed three regions of the intergalactic medium in the
interacting system HCG 92 corresponding to the regions observed in the
tidal dwarf galaxy survey of Lisenfeld \etal\ 2002 (their regions A,
B, and C). In addition, we have performed a 7-field mosaic the
``Arc-N'' region to the north of the tidal tail \citep[see Figure
\ref{schematic};][]{wil02}. The observations were taken over a period
of 2002 May 11 - 2002 Oct 19 using the 10 element
Berkeley-Illinois-Maryland Association (BIMA) millimeter
interferometer at Hat Creek California \citep{wel96}. Our observations
include 4 D array and 2 C array tracks on each region mentioned
above. We observed at the \twcooz\ spectral line (115.27 GHz) Doppler
shifted to a $V_{\rm LSR}$ given in column 4 Table \ref{pointings}
(optical definition) with a spectral raw resolution of 1.56 MHz, (4.06
\kms) with a total bandwidth of 386 GHz (960 \kms). When bright
emission was detected at region B, we followed up with more 2 C array
and 2 more D array tracks to increase signal to noise. The combined
data set includes baselines that range from 7 m to 88 m. The final
synthesized beam approximately 8\arcsec\ for each of the regions.

In addition to the the C- and D-array tracks for Region B, we obtained
two B-array tracks at a resolution of 2\arcsec. No emission was
detected in these data alone, and their inclusion into the combined
maps introduced artifacts and generally degraded the image
quality. Thus, the B-array data was not included in the final maps
shown here.

The CO emission for regions A, B, and C were believed to be relatively
centrally concentrated \citep{lis02}, so we observed these regions
with single pointings so as to not to use precious integration time
slewing. For the Arc-N region, we had no expectations of how extended
any CO emission may be, but there were reports of extended star
formation to the north of the tidal tail. We opted for the mosaic in
order to increase the area covered by our map. The primary beam of a
single point observation with the BIMA array at 115 GHz is 100\arcsec,
but with the 7-field mosaic, we are capable of covering a area with a
much flatter sensitivity in the inner 100\arcsec, and an effective
primary beam of around 180\arcsec. Details of the mosaic procedure
are given in \citet{hel03}.

Observing parameters for each pointing center are given in Table
\ref{pointings}. The first column is the label for each region we
observed. The second and third columns are the right ascension and
declination for the pointing centers of our observations. The fourth
column is the velocity to which we tuned, and the fifth column is the
number of pointings observed with these settings. The sixth, seventh
and eighths columns are the effective field of view, synthesized beam
(resolution) and r.m.s.~noise for each of the regions observed. Note
that all fields in this study are single pointings except for the
``Arc-N'' observations, which were done with a 7-field mosaic, and
thus have a higher noise level due to the overhead caused by slewing
during the track.

The data were calibrated and reduced using the MIRIAD data reduction
package \citep{sau95}. Instrumental and atmospheric phase variations
were measured by observing the quasar 2203+317 every 25
minutes. Amplitude calibration was performed using various planets as
primary flux calibrators. The data were weighted using the robust
weighting parameter of 0.5. The final channel maps have an
r.m.s. noise of $\sim 2-4 \times 10^{-2}$ \jybmkms.

\section{Morphology}

\subsection{Region A: The Intragroup Starburst \label{regionA}}

Previous observations of this region show CO emission that originates
from the center of the intragroup starburst (IGS). Unfortunately,
observations are complicated by the existence of two emission regions
separated in velocity space: one at around 6000 \kms, and the other at
around 6700 \kms. Despite the weakness of the CO emission at 6700
\kms\ observed by \citet{gao00}, the reality of their detection is
supported by the fact that it peaks near the center of the starburst
detected in the mid-IR and H-$\alpha$ maps of \citet{xu99} and is at
the same velocity of one of the \ion{H}{1} components \citep{wil02}.
The lower velocity CO emission was detected with the IRAM 30m by
\citet{lis02} and is reported to be more than twice as strong as the
higher velocity component. There is also star formation activity
observed at the lower velocity component as traced by the H$\alpha$
observations of \citet{xu99} (although it appears at around 5700 \kms,
not 6000 \kms\ like the CO detections).

Because the BIMA receivers cannot cover both emission regions with the
desired velocity resolution, we decided to observe the brighter
(previously unmapped at high resolution) low velocity component.
Despite reaching an r.m.s. noise of 21 m\jybm\ in the map (at 5 \kms\
resolution) we do not detect CO emission from this region. This is
somewhat surprising since the \ion{H}{1} column density and single
dish \twcooz\ fluxes of CO emission in Region A are actually
approximately equal or greater in intensity than the detections in
Region B \citep{wil02,lis02}, which we do detect. Thus the gas must
have a more even distribution rendering our interferometric
observations less sensitive to this more diffuse emission. We
note that with our tuning setup, we are unable to confirm the
existence of the IGM emission observed by \citet{gao00} since it lies
outside our velocity range at a higher velocity.

The IRAM 30m detection has a peak of $\sim$17 mK (T$_{\rm MB}$) in a
10 \kms\ channel which corresponds to a 75 mJy peak. Thus we would
have expected to see $>$3 sigma detection if the flux detected in the
IRAM beam were a point source. Conversely, if the flux detected by
IRAM were uniformly distributed across the 21\arcsec\ beam, we would
expect a flux of 8 mJy, which we would not have detected in our BIMA
map. If the flux detected by the single dish observations were
originating from two unresolved clumps, we would still predicted 1.5
sigma detections from these clouds. We can therefore conclude that the
molecular gas in the IGS is not made up of one small clump, but either
a more diffuse distribution, or at least 3 clumps separated by more
than about 8\arcsec\ (3.4 kpc at 88 Mpc).

\begin{figure}
\plotone{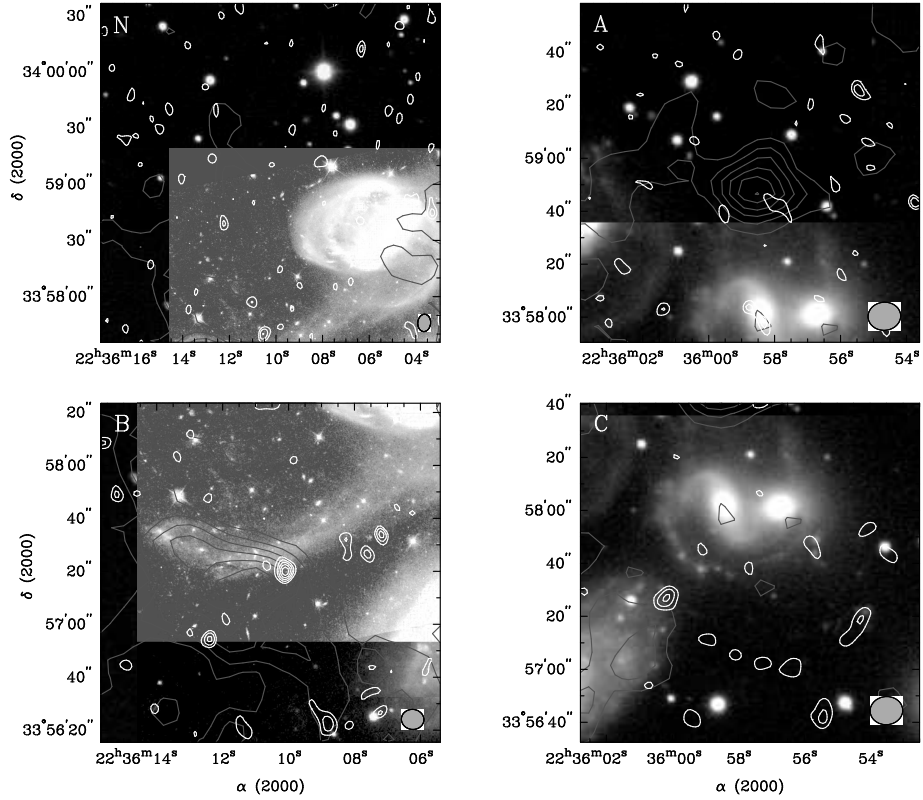}
\caption[f2.eps]{A raw moment zero map for the four regions
observed in this study. No clipping or masking has been applied in
order to show the noise levels reached in the regions where there are
no detections. The contour levels are 2,3,4... times the r.m.s. noise
in the moment map. For the three regions without a CO detection, we
integrate over 200 \kms around the \ion{H}{1} peak. For
Region N, the r.m.s. is 1.41 \jybmkms\ from 6500-6700 \kms; for
Region A the noise is 1.04 \jybmkms\ from 5900-6100 \kms; for Region
B the noise is 0.65 \jybmkms from 6500-6660 \kms; and for Region C
the noise is 0.64 \jybmkms\ from 6650-6850 \kms.
\label{fourregions}
}
\end{figure}

Figure \ref{fourregions} (upper left panel) shows the moment 0 map
overlain with the 2,3,4 sigma contours. The 3 sigma contours are
simply the result of spikes in the spectra that happen to lie within
the velocity range over which this moment map was made and do not
correspond to significant real emission. There is a clear lack of CO
emission at the peak of the \ion{H}{1} maps (overlaid with gray contours). One
possible explanation for this could lie in the existence of two
different velocity components at this region. We were only capable of
covering one within our tuning set up and choose the brightest of the
two, as seen in the single dish CO maps of \cite{lis02}. It is
possible that the component that is brighter in the single dish maps
(at 6000 \kms) contains more diffuse emission and perhaps the
component at 6700 \kms\ contains more dense clumps which were diluted
in the single dish maps. Follow up observations of this region tuned
to the 6700 \kms\ component would be useful in characterizing this
region in more detail.

\subsection{Region B: The Tidal Tail}

Previous studies of this region show the \ion{H}{1} peaks near the
intersection of the southern-most arc of the optical tidal tail with
the Arc-N portion of the \ion{H}{1} distribution. Previous single
dish \twcooz\ observations show strong emission from the optical tail
region \citep{lis02}. Follow-up Plateau de Bure interferometric
observations \citep{lis04} show this emission originates from the star
forming region detected by \citet{igl01} and lies on the dust lanes
seen in the HST images of \citet{gal01}. \citet{lis04} detect two
emission regions: one at the star forming dust lane (SQ B) and one at
the tip of the optical tidal tail (SQ tip). We do not detect CO at the
tip of the tidal tail, but it is a weak, marginal detection in the
more sensitive data of \citep{lis04}, so it is not unexpected that we
do not recover it with our less sensitive BIMA data.

\begin{figure}
\plotone{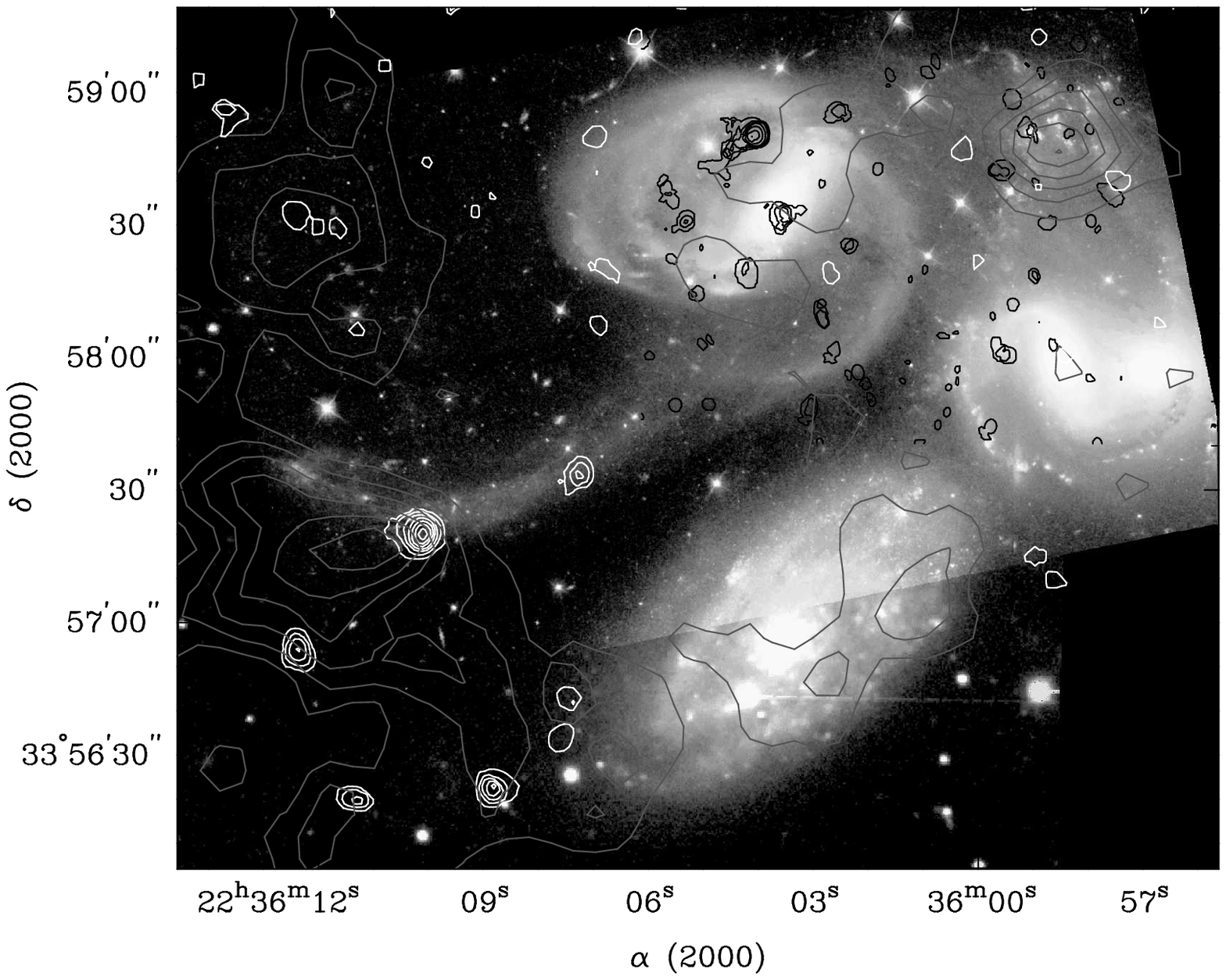}
\caption[f3.eps]{\twcooz\ contours overlaid on a composite image of
HCG 92. The white contours show the data obtained as part of this
study. The grey contours are the \ion{H}{1} observations of
\citet{wil02}. The black contours are the CO contours from the
\citet{gao00} study. The high resolution greyscale data is from HST
observations by \citet{gal01}; the lower resolution data used to fill
in the regions not covered by HST are taken from the Arp Catalog of
Peculiar Galaxies.  The contour levels for all maps are the same in
order to allow comparison of CO brightness between the regions with
detections.  The contour levels are 0.5, 1.0, 1.5 ... \jybmkms.  The
moment-0 CO maps in this figure and Figure \ref{moment0} used a
masking technique that does not show emission with a SNR $<$ 2. Thus,
the resulting map shows some features than are not visible in the
``raw'' map shown in Figure \ref{fourregions}.
\label{overview}
}
\end{figure}

\begin{figure}
\includegraphics[angle=0,scale=.7]{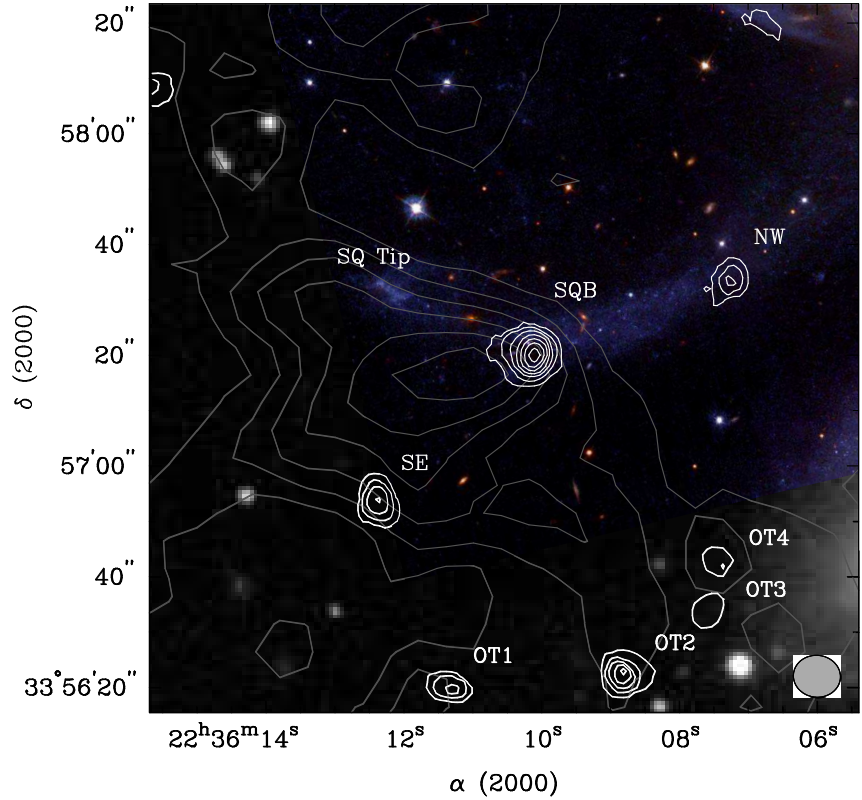}
%\plotone{f4.eps}
\caption[f4.eps]{A close up look at the CO emission detected
in the tidal tail region. The contour levels are 0.5, 1.0,
1.5...3.5 \jybmkms. The synthesized beam is shown as a black circle in
the lower right corner. The primary beam of this field is 100\arcsec,
indicating that the regions labeled OT1, OT2, OT3, and OT4 fall just
outside of the primary beam. Their spatial correspondence with the
``Old Tail'' and velocity coincidence with the \ion{H}{1} observations of this
region provides some evidence in support of the reality of these
features.
 \label{moment0}
}
\end{figure}

With our lower sensitivity, but larger field of view observations, we
detect three clumps of CO emission in the eastern-most tidal tail of
HCG 92 (Region B). Two of the clumps (SQB and SQB NW) lie directly on
the tidal tail and correspond well with dust lanes observed in the HST
images from \cite{gal01}. Figures \ref{overview} and \ref{moment0}
show these clumps overlaid on HST and DSS images, and the
correspondence with the dust lanes is obvious for the clumps that lie
on the optical tail. We note that the earlier CO observation of HCG 92
\citep[in particular NGC 7318,][]{yun97,gao00} show the CO emission in
the parent galaxy also corresponds well with the dusty regions visible
in the HST images \citep[][Figure \ref{overview}]{gal01}. The
correspondence of these regions of CO emission with the dust lanes is
not surprising. It is suspected that the same high density environment
that provide shielding against dissociation of dust molecules also
provides conditions conducive to the formation of molecules
\citep[\eg][and references therein]{alv99}.

{\it The detection of SQB NW is new with our observations -- the
Plateau de Bure interferometer maps of \citet{lis04} do not show it.}
Our detection of SQB NW falls outside the primary beam of the PdB
interferometer, thus the observations of \citet{lis04} were not
sensitive to this emission region. The same authors in \citet{lis02} did not
observe the location of SQB NW in their IRAM 30m mapping of HCG 92.
SQB NW does not have any corresponding \ion{H}{1} emission in the VLA map by
\citet{wil02}.

In addition to the two clumps in the tail, we detect a clump of
molecular gas to the south-east of the tail (SQB SE). This emission
region does not correspond with any previous observed star forming
regions or evident dust obscuration. Also, it seem to be at a slightly
different velocity than the other two regions in the tail (see Table
\ref{obstable}). Figure \ref{overview} shows the \ion{H}{1} contours
\citep{wil02} on the CO contours. While the CO emission at SQB SE does
not correspond with any features visible, it does lie within the \ion{H}{1}
cloud designated as ``Arc-N'' by \citet{wil02}. None of the peaks in
the CO emission we detect correspond with the \ion{H}{1} peak of the Arc-N
region. The two eastern-most peaks seem to straddle the \ion{H}{1} peak, while
the western CO peak does not correspond with any significant \ion{H}{1}
emission. \citet{lis02} pointed out that SQB does not lie at the peak
of the \ion{H}{1} emission, but rather at the location of a steep gradient in
\ion{H}{1} column density, and they suggest that perhaps compression of the
atomic gas at this location leads to formation of molecular gas. We
further point out that both the SQ tip detection by \citep{lis04} and
our new detection of SQB SE similarly lie on the edges of the peak in
\ion{H}{1} column density. The density gradient at the location of SQB is
larger than for SQ tip and SQB SE, which may explain why there is more
CO emission at these locations.

Figure \ref{spectra} shows the individual CO emission lines (solid)
with the \ion{H}{1} spectra overlaid with the dashed line in the ``New
Tail'' region.. The CO lines tend to lie within the envelope of the
\ion{H}{1} line, but skewed sightly toward the higher velocity
tail. This was not noted by \citet{lis04}, who find good agreement of
the centroid velocities of the \ion{H}{1} and CO in their higher
resolution maps. Comparing our spectra with those of
\citet{lis04} we find that our maps seem to be missing some emission
between the velocity range of 6576 - 6600 \kms. This is somewhat
surprising since our data are lower resolution and should be more
sensitive to more extended structure. It is likely that our
observations are less sensitive to this emission due to the larger
collecting area of the IRAM Plateau de Bure interferometer compared to
the BIMA array. Since the source is barely resolved for both
telescopes, spatial filtering is not as much of loss of signal as the
difference in collecting area.

The two extreme southern-most clumps lie in the middle of the ``Old
Tail.'' This feature appears as a faint optical structure that passes
behind the foreground galaxy NGC 7320 and has an \ion{H}{1} counterpart
\citep{wil02}. The origin of this tail is uncertain, and has recently
been ascribed to an old passage of NGC 7320c that pulled the tail out
of NGC 7318A \citep{wil02,sul01}. The same authors also ascribe the
more prominent tidal tail (which encompasses the CO detections SQB, SQ
tip and SQ NW) to a more recent passage of NGC 7320c that pulled a
tail out of NGC 7319.

\begin{figure}
\plotone{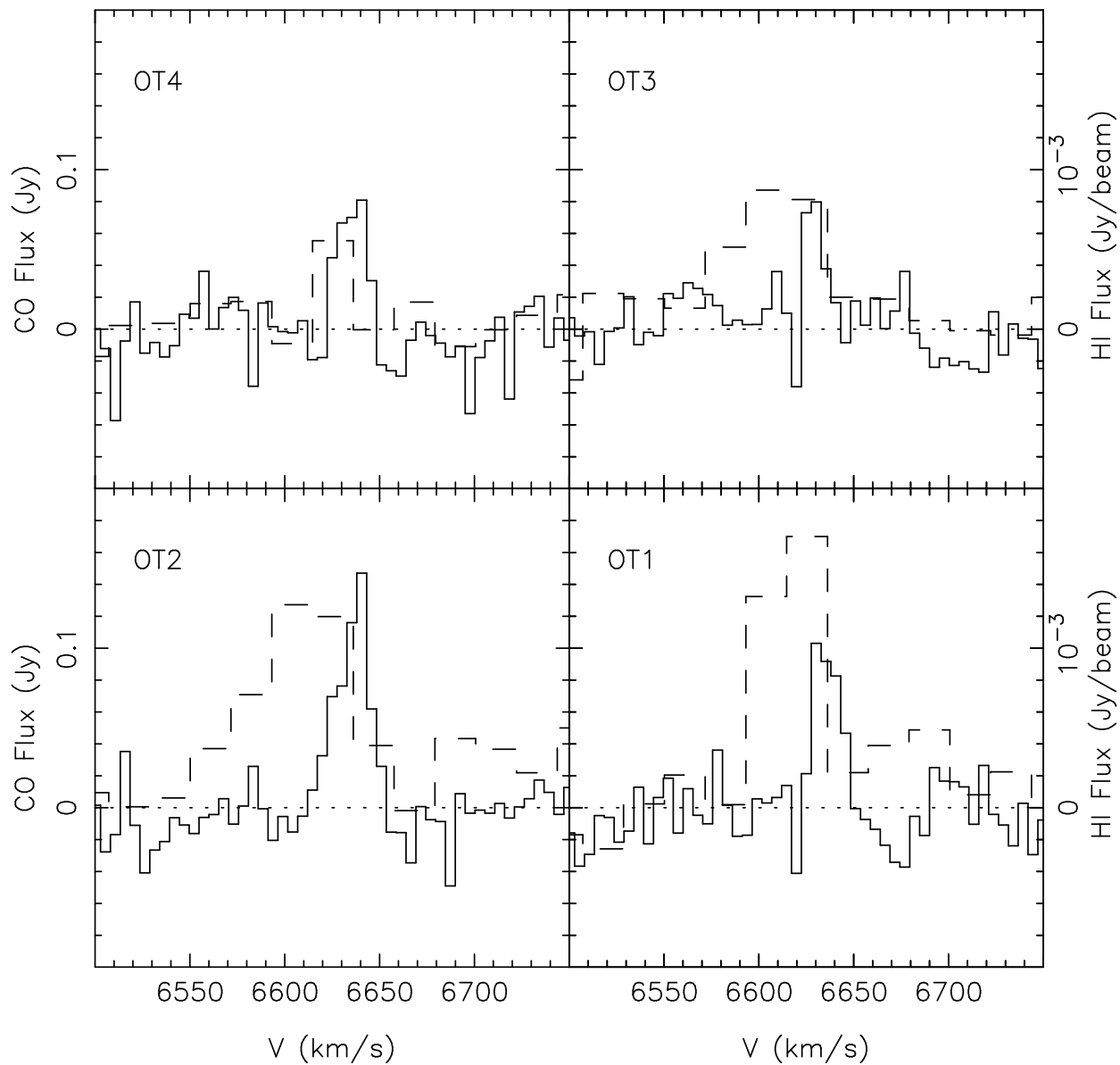}
\caption[f6.eps]{Individual \twcooz\ spectra for the four
emission regions discussed just outside the primary beam in the Old
Tail. The solid line is the CO data from this
study, and the dashed line is the \ion{H}{1} data from \citet{xu99}. The left
axis label corresponds to the CO and the right axis labels apply to
the \ion{H}{1} spectra.
\label{OTspectra}
}
\end{figure}

The two southern-most clumps OT1 and OT2 lie just outside the edge of
the BIMA primary beam. In an interferometer map, the sensitivity of
the telescope declines in an approximately Gaussian fashion from the
center to the edge. Normally objects appearing at the edge of the
primary beam (where the sensitivity has dropped to 50\% that of the
map center) would not be considered seriously. However, both OT1 and
OT2 show significant flux across 4 and 5 channels respectively, even
taking into account the reduced sensitivity (Figure \ref{OTspectra}).
Unfortunately the ongoing construction of the CARMA interferometer
from the antennae of BIMA and OVRO has made it impossible to obtain
new observations centered on the locations of OT1 and OT2.

If OT1 and OT2 are real, they might be molecular gas associated with
the ``Old Tail'', which they overlap spatially. \citet{sul01} use
various estimators to derive an age of this tail between 6 and
12$~\times~10^8$ yr, compared to 2 to 4$~\times~10^8$ yr for the more
visually prominent younger tail. If OT1 and OT2 are ``Old Tail''
clouds, they may have existed in intergalactic space for $\sim$
10$^8$ yr. Alternately they may have formed more recently, triggered
by the interaction of NGC 7320c with the gas in the ``Old Tail''
during its second passage through the group.

OT1 and OT2 also overlap spatially with a region of far UV emission
detected by {\it GALEX} which yielded an age of $\sim 3~\times~10^8$
yr for the stellar populations and a current star formation rate of
0.060 $\pm$ 0.013 \msun yr$^{-1}$ \citep{xu05}.

In the Old Tail region, the velocity offset between the CO and
\ion{H}{1} emission is similar to the offset seen in the ``New
Tail'' (see Figure \ref{OTspectra} and Table \ref{OTspectra}). {The
offset appears to be slightly larger in the ``Old Tail'' region
compared to the ``New Tail'' emission since} in these regions, we do
not see any high velocity tail and do not have the signal to noise to
detect any such tail in our lines. 

%Thus we feel that the offset seen
%between the CO peak and the \ion{H}{1} peak is the Old Tail region is
%real.

\subsection{SQ C: Star Forming Tidal Dwarf Galaxies?}

Region C in Figure \ref{overview} lies to the west of Region B, and is
the site of a number of candidate TDGs listed by \citep{men01}, who
isolated them based upon H$\alpha$ emission. \citep{lis02} did not
detect CO emission from any of these candidates in IRAM 30m
observations targeted at them and at a few nearby local \ion{H}{1} maxima.
Our single BIMA pointing covers the entire region and places an upper
limit of 25 m\jybm. Although there are indeed a number of star
clusters in this area coincident with the H$\alpha$ emitting regions
(see the analysis of the data of \citet{gal01} by \citet{lis02}), in
the absence of any kinematically distinct component of ISM, no strong
case can be made for the presence of TDGs in the western side of HCG
92.

\subsection{Arc-N: Star Formation Without an Optical Tail?}

The Arc-N region \citep{wil02} is an \ion{H}{1} rich gaseous tail that is
likely the result of the gas in the parent galaxies (NGC 7319, NGC
7318a,b) being stripped out by the rapid passage of NGC 7320c through
the system. Recent GALEX images show UV emission originating from
various regions within the Arc-N area \citep{xu05}. This the strength
of this emission is indicative of star formation rates on the order of
0.3 \msun\ yr$^{-1}$.

Again, despite the evidence of ongoing star formation and an abundance
of \ion{H}{1} emission, we do not detect any significant CO emission in this
region. Our 7-point mosaic covered all the \ion{H}{1} emission \citep{wil02}
and star formation regions detected by GALEX \citep{xu05} and reached
a sensitivity of 39 m\jybm. We discuss the mass detections limits in
the next section.

\section{Fluxes and Mass Estimates}

\subsection{Region B}

We detect three emission regions within the BIMA primary beam around
the tidal tail in HCG 92. The flux for each individual clump in
region B is shown in Table \ref{obstable}. The main CO peak on the
tidal tail has also been recently detected by \citet{lis04} using the
IRAM interferometer. They detect a flux of 4.4 $\pm$ 1.0 \jykms\ for
this region, which agrees with our value of 3.8 $\pm$ 0.4 \jykms. The
IRAM 30-m detects 5.2 \jykms \citep[][assuming a Jy-K conversion
factor of 4.7]{lis02}. Thus, the flux we detect with BIMA is
approximately 70\% of the single dish flux, suggesting that 30\% of
the molecular gas structures are on scales larger than approximately
8\arcsec.

\citet{lis04} also detect weak emission at the tip of the tidal tail,
which is too faint to be detected in our data. Additionally, the IRAM
30-m maps to not extend far enough to cover the two weaker emission
peaks in the tidal tail region (SQB NW and SQB SE), so we cannot
determine what spatial scales dominate the molecular gas distribution.

To calculate the mass for each clump, we summed the flux over a box
that covered the emitting regions. The size of the boxes used is given
in column 4 of Table \ref{obstable}. The mass is calculated using the
equation
$$ M_{\rm mol} ({\rm M_\odot}) = 1.61 \times
10^4~\left({\alpha\over{\alpha_{\rm Gal}}} \right) ~ d^2_{\rm Mpc} ~
S_{\rm CO} \eqno{(1)}$$
where $\alpha/\alpha_{\rm Gal}$ is the CO-to-H$_2$ conversion factor
relative to the Galactic value ($\alpha_{\rm Gal} = 3\pm 1 \times
10^{20}~{\rm cm}^{-2} ({\rm K~km~s}^{-1})^{-1}$; \citet{sco87}),
$d_{\rm Mpc}$ is the distance to Stephan's Quintet in Mpc, and
$S_{\rm CO}$ is the \twcooz\ flux in \jykms. A factor of 1.36 is
included to account for the presence of helium.

Assuming a distance of 88 Mpc (for H$_0$ = 75 \kms/Mpc) we obtain a
molecular mass for the central CO peak of 4.7 $\times 10^8$ \msun\
assuming a Galactic CO-to-H$_2$ conversion factor. If the molecular
gas in SQB is similar in its properties to the giant molecular clouds
in the Milky Way then we can assume that most of the mass is in clouds
$\sim$ 10$^6$ \msun. In this case, SQB likely consists of $\sim$
10$^2$ clouds of that mass, which are the precursors to the formation
of OB associations. In fact, the H$\alpha$ imaging of HCG 92 by
\citet{igl01} shows that the location of SQB is the only spot in the
tidal arm with any H$\alpha$ emission.

\subsection{Regions A, C, and N}

We did not detect any CO emission in our observations of either region
A, C, or N. In region A our observations reached a noise level of 21
m\jybm, while in Region C we reached a noise level of 25 m\jybm.
Assuming a minimal real detection would be spatially unresolved, with
velocity width equal to three continuous velocity channels, we would
be unable to detect features more massive than 3.9 $\times 10^{7}$
\msun in Region A, and 4.6 $\times 10^{7}$ \msun in Region C.

The r.m.s noise reached in the Arc-N region was higher because the
sensitivity was spread over 7 fields instead of concentrating on a
single field as we did in the other regions. The final maps cover an
effective area of 180\arcsec\ and reach an r.m.s. of 39 m\jybm. At a
distance of 88 Mpc, this corresponds to an upper mass of 7.3 $\times
10^7$ \msun.

At these mass limits, we can rule out supergiant molecular clouds
such as those observed in the Antennae \citep{wil03} as the source of
the star formation since the supergiant clouds generally have masses
of a $10^8$ \msun. It thus seems most likely that regular giant molecular
clouds are the source of the ongoing star formation observed in these
three regions.

As discussed in \S\ref{regionA} our lack of detections in region SQA
despite the strong flux detected in this region with the IRAM 30m
\citep{lis02} suggests that the molecular gas in SQA is either
diffuse or exists in more than two clumps separated by more than 3.4
kpc.

\section{Discussion}

\subsection{Are these Clouds Self Gravitating?}

To determine if this object is gravitationally bound, we can compare
the mass determined by the flux with what we would expect from the
linewidth if the object is in virial equilibrium. If the object is
virialized, then its mass can be determined from the following equation:
$$
M_{\rm vir} ({\rm M_\odot}) = 99~V^2_{\rm FWHM}~D_{\rm pc}
\eqno{(2)}
$$
where $V_{\rm FWHM}$ is the full width half-maximum of the spectral
line, and $D$ is the cloud diameter (given by 0.7($D_\alpha +
D_\delta$), where $D_\alpha$ and $D_\delta$ are the molecular
clouds diameters along the east-west and north-south directions
respectively in parsecs; see \citet{wil95}) and a cloud with a density profile
that varies as $r^{3/2}$. Assuming a marginally resolved cloud in our
map, we obtain a diameter of 4800 pc. A Gaussian fit to the line
profile gives a FWHM of 20.3 \kms, yielding a virial mass of $2.0
\times 10^8$ \msun. For the other two clouds, we assume the same
diameter (unresolved) but adopt the FWHM line widths given in Table
\ref{obstable} to calculate the virial mass. The results are shown in
the last column of Table \ref{obstable}.

Since the molecular mass of SQB determined from the CO flux is greater
than the mass derived from the linewidth for a virialized cloud, it
is likely that this object is massive enough to be self-gravitating.
Thus, this object likely represents a long-lived clump that will
survive \citep[given its current star formation rate of 0.3 - 0.5
 \msun year$^{-1}$][]{xu05,lis04} should survive for at least a few
$\times 10^8$ years.

One complication to this simple picture has been raised by 
\citet{hb04} who argue that the kinematics observed in TDGs 
may not necessarily arise from the gravitational potential of
the TDGs, but rather from the flow of gas along tidal arms
as a result of the galaxy-galaxy interactions. In this view,
the dense concentrations of gas seen in TDGs may be a chance
alignment of our line of sight {\it along} a segment of a tidal
arm, yielding a large column density of gas, but without a large 
volume density. Although these authors cast doubt on the reality
of many TDGs, they concede that such objects are expected to form
in tidal arms. \citet{hb04} argue that best possible situation in
which to determine the dynamical mass (and hence reality) of a 
TDG is when the viewing geometry is face-on to the tidal arms.

In the case of Stephan's Quintet, the geometry is difficult to 
determine. Comparing with previously published interaction models,
\citet{sul01} suggest that the specific interaction that created
the more prominent, younger tidal tail coming from NGC 7319 was
a low velocity one, with a low inclination with respect to the 
the disk of NGC 7319. Unfortunately, the disk of NGC 7319 is gas 
poor and no kinematic estimates of its inclination are available 
in the literature. The visible appearance of NGC 7319 and its
tidal arm suggest that both lie close to the plane of the sky --
nearly face on. Our estimates of the dynamical masses are 
likely at the lowest possible risk for being affected by poor 
viewing geometry.

The strong CO flux suggests that the molecular gas in this cloud is
optically thin (which is supported by the high \twco/\thco\ ratio
reported by \citet{lis04}). At first glance this is not surprising -- the
first gas to be stripped off during an encounter should be the outer
gas, which in spiral galaxies is most often at lower metallicities.
However, spectroscopy by \citet{lis04} has shown the oxygen abundance
(12 + log O/H) to be 8.7, nearly equal to the solar value. This is
high for the outer regions of typical spiral galaxies, and is high
for other TDGs \citep{duc04}. Since HCG 92 is believed by many
authors to have been the site of repeated encounters, it is possible
that by the time the ``Young Tail'' was formed all the gas from the
outer regions had already been stripped away. The gas we observe in
SQB may therefore be from the interior of NGC 7319.

The two clouds candidates identified in the `` Old Tail'', OT1 and
OT2, if real, are more likely to be from the outer regions of their
parent galaxy.

If the gas is optically thin, but a low metal abundance is not the
reason, it is likely this state is caused by the influence on the ISM
of the prominent star forming region in SQB. This idea was proposed
by \citet{lis04} to explain the high $^{12}$CO / $^{13}$CO ratio they
observed. It is consistent with recent estimates of the star
formation rate, which range from 0.5 \msun yr$^{-1}$ \citep{lis04} to
0.3 \msun yr$^{-1}$ \citep{xu05}.

For some nearby starburst galaxies it has been argued that the low
opacity of the CO emission could cause a reduction in the CO-to-H$_2$
conversion factor ($\alpha/\alpha_{\rm Gal}$) (\eg\ \citet{hhr99}
among others). If true this means we are overestimating the
molecular mass. This would lead to our molecular mass being an
upper limit, and would make our arguments about the gravitationally
bound nature of SQB weaker.

\subsection{CO-H$_2$ Conversion Factor Limits}

Adopting the method used by \citet{wil95} we can estimate the
CO-to-H$_2$ conversion factor ($\alpha/\alpha_{\rm Gal}$) by assuming
the clouds are virialized, then seeing what value of
$\alpha/\alpha_{\rm Gal}$ is needed to equate the virial mass to the
molecular mass. Since the clouds in this study are not resolved, we
will only be able to place constraints on $\alpha/\alpha_{\rm Gal}$.

For unresolved clouds, our virial mass will be an over-estimated. Our
molecular mass will also likely be an overestimate, but to a lesser
degree. To calculate $\alpha/\alpha_{\rm Gal}$, we set equation number
(1) equal to equation number (2) and solve for $\alpha/\alpha_{\rm
Gal}$. More simply put, $\alpha/\alpha_{\rm Gal}$ = M$_{\rm
vir}/$M$_{\rm mol,Gal}$ where M$_{\rm mol,Gal}$ is the molecular mass
determined assuming a Galactic value for $\alpha/\alpha_{\rm
Gal}$. This yields values of $\alpha/\alpha_{\rm Gal}$ of 0.4, $<$ 0.4
and $<$ 1 for the clouds SQB, SQB SE and SQB NW, respectively.
Despite the near solar metal abundance, there is at lease some
evidence evidence for a non-Galactic value of $\alpha/\alpha_{\rm
Gal}$ in CO emission in the tidal arms of HCG 92.

\section{Summary}

We present the first complete high resolution study of molecular gas
in Stephan's Quintet. Our observations are centered on all major
overlap regions in this interacting system in order to study the Tidal
Dwarf Galaxy population and the conditions resulting in the
intracluster star formation activity observed at other wavelengths.

The wide field of view of the BIMA array enable to detect 7 separate
clumps of molecular gas in the new and old tidal tails to the east of
Stephan's Quintet. One of these clumps has already been studied with
the Plateau de Bure interferometer by \citet{lis04} but the remaining
6 clumps are new detections. Of these 6, two are well within the
primary beam, and likely correspond with the bright ``New'' tidal tail to the
east of Stephan's Quintet. The remaining four detections are tentative
detections at or just beyond the primary beam and lie coincident
(spatially and in velocity) with the ``Old Tail'' region of Stephan's
Quintet.

All but one of the CO clumps associated with the new and old tails
correspond well with the velocity of and distribution of the \ion{H}{1}
observation of \citet{wil02}. They also correspond well spatially with
the evidence of on-going star formation discovered by recent GALEX
observations \citep{xu05}. One of the clumps in the new tail (SQB NW)
however does not seem to have any corresponding \ion{H}{1} emission,
but there is evidence of star formation seen in UV images of this
region. The nature of this region is a bit of a
mystery and will need to be studied further at other wavelengths.

Adopting a Galactic CO-to-H$_2$ conversion factor, we estimate the
molecular masses for these clumps to be on the order of $ 1 - 5 \times
10^8$ \msun. The linewidths for these regions suggests virial masses
of approximately 50\% of the molecular mass, suggesting that these
clouds are gravitationally bound, non-transient objects. Our data
shows some evidence that the Galactic value of the CO-to-H$_2$
conversion factor is not appropriate in this near solar metallicty
region.

In all other regions we did not detect any significant CO
emission. This is somewhat surprising since the IRAM 30-m observations
by \citet{lis02} showed strong CO emission in the intracluster
starburst region (SQ A). Our observation of SQ A reached a sensitivity
of 21 m\jybm\ (at 5 \kms\ velocity resolution) corresponding to a mass
sensitivity of $3.9 \times 10^7$ \msun\ (for an unresolved source).
Our lack of detection suggests the molecular ISM in this starburst
either consists of a diffuse gas or at least 3 clumps separated by
over 3 kpc.

\acknowledgments 

G.~R.~P.~is supported in part by NSF grant AST-0228974 and by the
State of Maryland via support of the Laboratory for Millimeter-Wave
Astronomy. We thank the referee for the careful comments that greatly
improved the quality of this paper. The authors are grateful to
Lourdes Verdes-Montenegro for providing the \ion{H}{1} data cube of HCG 92.
We also thank Nathaniel Taylor for waiting until this paper was
submitted, allowing an extra week of Chris' attention, and Melissa for
her patience.

%\clearpage

\clearpage

\begin{figure}
\plotone{f5.eps}
\caption[f5.eps]{Individual \twcooz\ and \ion{H}{1} spectra for the four
regions discussed in the text. The solid line is the CO data from this
study, and the dashed line is the \ion{H}{1} data from \citet{wil02}. The left
axis label corresponds to the CO and the right axis labels apply to
the \ion{H}{1} spectra. The upper left panel is the non-detection
of the ``SQ Tip'' region shown in \citet{lis04}; the upper right panel is the
weaker emission region to the north-west of the tidal tail in the IGM
(SQB NW); the lower left panel is the peak in the south-east of the
tidal tail (SQB SE); and the lower right panel is the central
emission peak on the tidal tail (SQB). Note, that there {\it is}
significant \ion{H}{1} emission in SQB SE, it is just scaled down
in order to be able to show the \ion{H}{1} emission in region SQB and
SQ Tip with the same vertical scale.
\label{spectra}
}
\end{figure}

\clearpage

\begin{deluxetable}{lcccccccr}
\tablecolumns{9}
\tablewidth{0pc}
\tabletypesize{\small}
\tablecaption{Observing Parameters \label{pointings}}
\tablehead{
\colhead{label} & \colhead{RA}     & \colhead{Dec.}   &
\colhead{$V_{\rm LSR}$} & \colhead{pntgs} & \colhead{f.o.v}  & \colhead{synth.~beam}     & \colhead{r.m.s @ 5\kms}    \\
\colhead{~}     & \colhead{(J2000)}& \colhead{(J2000)}& \colhead{(\kms)}   & \colhead{~}         & \colhead{(\arcsec)}    & \colhead{\arcsec$\times$\arcsec} & \colhead{(m\jybm)} \\
          } 
\startdata
% label&  RA            &   dec                      & Vlsr  &pntgs& prim.bm &   synth.beam     &  rms  
SQA    & 22\h35\m58.7\s &33\arcdeg58\arcmin55\arcsec & 6050  &  1  &   100   &  7.1$\times$6.3  & 21   \\
SQB    & 22\h36\m10.5\s &33\arcdeg57\arcmin20\arcsec & 6625  &  1  &   100   &  8.5$\times$7.5  & 23   \\
SQC    & 22\h35\m57.6\s &33\arcdeg57\arcmin37\arcsec & 6600  &  1  &   100   & 12.2$\times$10.6 & 25   \\
Arc-N  & 22\h36\m12.0\s &33\arcdeg58\arcmin28\arcsec & 6625  &  7  &   180   &  9.3$\times$6.8  & 39   \\
\enddata
\end{deluxetable}

\begin{deluxetable}{lcccccccr}
\tablecolumns{9}
\tablewidth{0pc}
\tabletypesize{\small}
\tablecaption{New Tail Line Parameters \label{obstable}}
\tablehead{
\colhead{label} & \colhead{RA}     & \colhead{Dec.}  & \colhead{area}          & \colhead{flux}    & \colhead{FWHM} & \colhead{V}    &\colhead{M$_{\rm gas}$}   &\colhead{M$_{\rm vir}$}   \\
\colhead{~} & \colhead{(J2000)}&\colhead{(J2000)}&\colhead{($''\times ''$)}&\colhead{(\jykms)}&\colhead{(\kms)}&\colhead{(\kms)}&\colhead{(\msun)}        &\colhead{(\msun)}         \\
          } 
\startdata
SQB    & 22\h36\m10.1\s &33\arcdeg57\arcmin20\arcsec & 13$\times$11 & 3.8$\pm$0.4 & 20.3 & 6635 & 4.7 $\times 10^8$ &     2.0 $\times 10^8$ \\
SQB SE & 22\h36\m12.3\s &33\arcdeg56\arcmin53\arcsec & 6$\times$10  & 2.2$\pm$0.4 & 16.1 & 6614 & 2.7 $\times 10^8$ & $<$ 1.2 $\times 10^8$ \\
SQB NW & 22\h36\m07.3\s &33\arcdeg57\arcmin34\arcsec & 8$\times$8   & 1.9$\pm$0.4 & 22.2 & 6632 & 2.4 $\times 10^8$ & $<$ 2.3 $\times 10^8$ \\
\enddata

\end{deluxetable}

\begin{deluxetable}{lcccccccr}
\tablecolumns{9}
\tablewidth{0pc}
\tabletypesize{\small}
\tablecaption{Old Tail Line Parameters \label{OTtable}}
\tablehead{
\colhead{label} & \colhead{RA}     & \colhead{Dec.}  & \colhead{area}          & \colhead{flux}    & \colhead{FWHM} & \colhead{V}    &\colhead{M$_{\rm gas}$}   &\colhead{M$_{\rm vir}$}   \\
\colhead{~} & \colhead{(J2000)}&\colhead{(J2000)}&\colhead{($''\times ''$)}&\colhead{(\jykms)}&\colhead{(\kms)}&\colhead{(\kms)}&\colhead{(\msun)}        &\colhead{(\msun)}         \\
          } 
\startdata
SQOT 1 & 22\h36\m11.3\s &33\arcdeg56\arcmin20\arcsec & 14$\times$10  & 1.8$\pm$0.3 & 15.2 & 6636 & 2.2 $\times 10^8$ & $<$ 1.1 $\times 10^8$ \\
SQOT 2 & 22\h36\m08.8\s &33\arcdeg56\arcmin23\arcsec & 14$\times$12  & 2.8$\pm$0.4 & 21.0 & 6636 & 3.5 $\times 10^8$ & $<$ 2.1 $\times 10^8$ \\
SQOT 3 & 22\h36\m07.6\s &33\arcdeg56\arcmin34\arcsec & 10$\times$10  & 0.9$\pm$0.3 &  9.9 & 6629 & 1.1 $\times 10^8$ & $<$ 0.5 $\times 10^8$ \\
SQOT 4 & 22\h36\m07.4\s &33\arcdeg56\arcmin43\arcsec &  8$\times$8   & 1.4$\pm$0.3 & 16.4 & 6635 & 1.7 $\times 10^8$ & $<$ 1.3 $\times 10^8$ \\
\enddata

\end{deluxetable}

\end{document}